# Helicity of the toroidal vortex with swirl

E. Yu. Bannikova[1,2], V.M. Kontorovich[1,2] and S.A. Poslavsky[2]

[1] Institute of Radio Astronomy, National Academy of Sciences of Ukraine
Krasnoznamennaya St. 4, Kharkov, 61002 Ukraine

[2] Karazin Kharkov National University
Svobody Sq. 4, Kharkov, 61022 Ukraine

On the basis of solutions of the Bragg-Hawthorne equations we discuss the helicity of thin toroidal vortices with the swirl – the orbital motion along the torus directrix. It is shown that relationship of the helicity with circulations along the small and large linked circles – directrix and generatrix of the torus - depends on distribution of the azimuthal velocity in the core of the swirling vortex ring. In the case of non-homogeneous swirl this relationship differs from the well-known Moffat relationship – the doubled product of such circulations multiplied by the number of links. The results can be applied to vortices in planetary atmospheres and to vortex movements in the vicinity of active galactic nuclei.

## 1. Introduction

In nature, the toroidal vortices often have a 'swirl' [1] – the orbital motion along the torus directrix. Such objects are the attached ring vortices of tropical cyclones, hurricanes and tornadoes [2], as well as solar toroidal vortices [3] responsible for the 11-year cycle of activity, and many others (see, e.g. [4]). In the presence of swirl there appears a topological integral - the helicity [5]. Laboratory experiments have confirmed that this additional integral may increase the vortex stability [6].

It is known that for two linked vortex contours the helicity should be equal to the product of the circulations multiplied by the doubled number of links [1, 5, 7, 8]. We will present the example with hydrodynamic solution showing that for the toroidal vortex with swirl this ratio has some different form, reflecting spatial distribution of the vorticity.

## 2. Bragg-Hawthorne equation and its solution for toroidal vortex

Let us consider the axisymmetric stationary flow of an ideal incompressible fluid in the absence of external forces. Euler's equation in this case has the form $(\mathbf{V}\nabla)\mathbf{V} = -\nabla p/\rho$. Taking into account $(\mathbf{V}\nabla)\mathbf{V} = grad\,(V^2/2) - \mathbf{V} \times rot\mathbf{V}$, we rewrite it as follows:

$$\mathbf{V} \times rot\mathbf{V} = \nabla\left(\frac{p}{\rho} + \frac{V^2}{2}\right), \tag{2.1}$$

where in cylindrical coordinates $(r, \varphi, z)$

$$\omega = rot\mathbf{V} = \left(-\frac{\partial V_\varphi}{\partial z}\mathbf{i}_r, \left(\frac{\partial V_r}{\partial z} - \frac{\partial V_z}{\partial r}\right)\mathbf{i}_\varphi, \frac{1}{r}\frac{\partial (rV_\varphi)}{\partial r}\mathbf{i}_z\right). \quad (2.2)$$

We will use the Stokes stream function $\psi$ defined in the cylindrical coordinates $(r, \varphi, z)$ in accordance with

$$V_r = -\frac{1}{r}\frac{\partial \psi}{\partial z}, \quad V_z = \frac{1}{r}\frac{\partial \psi}{\partial r}. \quad (2.3)$$

Due to (2.3) the continuity equation $div\mathbf{V} = 0$ is identically satisfied. Orbital velocity component can now be written in the following form

$$V_\varphi = \frac{f(\psi)}{r}, \quad (2.4)$$

where $f(\psi)$ is an orbitrary function. Substituting the expressions (2.3) and (2.4) into (2.2), we obtain

$$rot\mathbf{V} = \left(-\frac{1}{r}f'\frac{\partial \psi}{\partial z}\mathbf{i}_r, -\frac{\tilde{\Delta}\psi}{r}\mathbf{i}_\varphi, \frac{1}{r}f'\frac{\partial \psi}{\partial r}\mathbf{i}_z\right) \quad (2.5)$$

where

$$\tilde{\Delta}\psi \equiv r\frac{\partial}{\partial r}\left(\frac{1}{r}\frac{\partial \psi}{\partial r}\right) + \frac{\partial^2 \psi}{\partial z^2}, \quad f' \equiv \frac{df}{d\psi}.$$

Substituting (2.5) into the Euler equation (2.1) after simple transformations we obtain for the $\varphi$-th component of the vorticity:

$$rot_\varphi \mathbf{V} = -\frac{f \cdot f'}{r} + \frac{r}{\psi'_r}\frac{\partial \Pi}{\partial r}, \quad (2.6)$$

where $\Pi = p/\rho + V^2/2$ is the Bernoulli integral. Equating the expressions (2.5) and (2.6) for the azimuthal component of the vorticity, we obtain the Bragg-Hawthorne equation for $\psi$ [1, 9] (or, equivalently, the Grad-Shafranov equation in MHD case [10]) with given $\Pi$ and $f$:

$$\frac{\partial^2 \psi}{\partial r^2} - \frac{1}{r}\frac{\partial \psi}{\partial r} + \frac{\partial^2 \psi}{\partial z^2} = r^2\frac{d\Pi}{d\psi} - f\frac{df}{d\psi}. \quad (2.7)$$

Solutions of the equation (2.7) for $f \neq 0$ describe the stationary axisymmetric flows with swirl. The set of such solutions can be obtained for the case (that conserves the equation as linear)

$$\frac{d\Pi}{d\psi} = Const = \alpha\psi_0; \quad f\frac{df}{d\psi} = Const = -\beta R^2\psi_0 , \tag{2.8}$$

where $\alpha, \beta$ are dimensionless constants and $\psi_0$ is dimensional normalization factor. Then the solution of equation (2.7) for the function of Stokes flow has the form [10][1]:

$$\psi = \psi_0\left[\frac{1}{2}\left(\beta R^2 + r^2\right)z^2 + \frac{\alpha-1}{8}\left(r^2 - R^2\right)^2\right]. \tag{2.9}$$

Using it, we obtain $V_r$ и $V_z$ from (2.3), and using (2.8) we assign the swirl velocity $V_\varphi$ (and pressure $p$). Close to the circle $r = R; z = 0$ (i.e. with $|r-R|/R \ll 1$ and small $z$) the expression (2.9) for the flow function becomes

$$\psi \simeq \frac{1}{2}\psi_0 R^2\left[(\beta+1)z^2 + (\alpha-1)(r-R)^2\right]. \tag{2.10}$$

In this solution, with $\beta+1>0$ and $\alpha>1$, surfaces $\psi(r,z)=Const$ are the nested tori with the common rotary axis (directrix) $r = R, z = 0$, and their meridional section are ellipses. From the equations (2.8) we obtain:

$$\Pi = \alpha\psi_0\psi + \Pi_0 \qquad f^2 = f_0^2 - 2\beta R^2\psi_0\psi , \tag{2.11}$$

where $\Pi_0$ and $f_0$ are integration constants parametrizing the solution.

## 3. Helicity

In the next sections we will find the helicity integral [5] for the solution (2.10)

$$S = \int \mathbf{V}\cdot rot\mathbf{V}dV . \tag{3.1}$$

We restrict ourselves to a particular case $\beta+1 = \alpha-1$, when sections of the stream surface by meridional plane are the circles:

$$\psi \simeq \frac{1}{2}\phi_0 R^2(\alpha-1)\left[z^2 + (r-R)^2\right]. \tag{3.2}$$

---

[1] Other examples of the solutions, mainly in connection with the study of MHD configurations, can be found in the monograph [11] and in the collections of reviews "Questions of plasma theory" edited by M.A. Leontovich (Moscow, Atomizdat, 1963-1982)

One of the tori $\psi(r,z) = Const$ can be considered as a boundary of the area occupied by the **swirling vortex flow**. We assume that meridional cross-section of the boundary is the circle with radius $a$ (Fig. 1), and the flow outside this torus is potential.

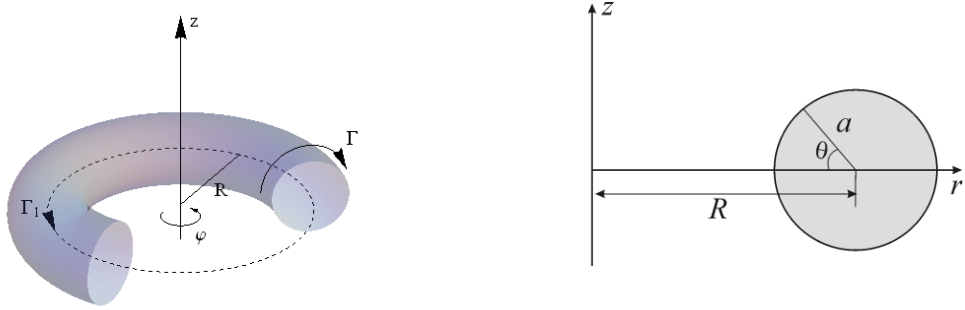

Fig.1 Left: 3D-scheme of toroidal vortex with swirl; right – section in a meridian plane.

Taking into account (2.3), we obtain expressions for the velocity components:

$$V_r = -R^2 \psi_0 (\alpha - 1) \frac{z}{r}, \quad V_z = R^2 \psi_0 (\alpha - 1) \frac{r - R}{r}. \tag{3.3}$$

In the case of a thin ring vortex this movement in the meridional plane has the nature of 'solid-body rotation', as the rotation linear velocity is proportional to the distance to the circular directrix of the vortex ring:

$$V_r \simeq -R \psi_0 (\alpha - 1) z, \quad V_z \simeq R \psi_0 (\alpha - 1)(r - R).$$

Here the azimuthal component of the vorticity in the thin ring is practically constant:

$$\omega_\varphi \simeq -2R \psi_0 (\alpha - 1).$$

## 4. The case of homogeneous swirl of vortex ring

Firstly, we will consider the case $\beta = 0$, $\alpha = 2$ for which $f(\psi) \equiv f_0 = Const$. This means that the swirl inside the torus is distributed homogeneously: $V_\varphi = f_0/r \simeq f_0/R$. Two other velocity components and azimuthal component of the rotor can be written as

$$V_r \simeq -R \psi_0 z, \quad V_z \simeq R \psi_0 (r - R),$$
$$\omega_\varphi \simeq -2R \psi_0. \tag{4.1}$$

It is convenient to pass to the polar coordinate system in the meridional plane with origin at the point $r = R$, $z = 0$: $R - r = \eta \cos \theta$, $z = \eta \sin \theta$ ($\eta$ is the coordinate measured from the rotary axis of the torus along the radius of its cross-section). At the boundary $\eta = a$ we have the vortex sheet $\omega_\theta = V_\varphi \delta(\eta - a)$ because the azimuthal velocity is discontinuous here (we assume the velocity equals zero outside the torus)[2].

In the presence of a swirl inside the torus – (orbital motion with angle $\varphi$) and a vortex motion in meridional section along the small contour (with angle $\theta$) – it is convenient to represent the helicity as a sum of two terms

$$S = S_\varphi + S_\theta, \qquad S_\varphi = \int V_\varphi \omega_\varphi dV, \qquad S_\theta = \int (V_r \omega_r + V_z \omega_z) dV. \qquad (4.2)$$

The components $S_\varphi$, $S_\theta$ of the helicity (4.2), with (4.1) and $V_\theta^2 = V_r^2 + V_z^2$ taken into account, have the following form:

$$S_\varphi = \int V_\varphi \omega_\varphi dV \simeq -4\pi^2 R a^2 f_0 \psi_0,$$

$$S_\theta = \int (V_r \omega_r + V_z \omega_z) dV = \int V_\theta \omega_\theta dV = \int_\Sigma V_\theta V_\varphi dS. \qquad (4.3).$$

In the last expression we integrate by the surface $\Sigma$ of the vortex ring. Since on this surface $V_\theta \simeq -Ra\psi_0$, $V_\varphi \simeq f_0/R$, then $S_\theta \simeq -4\pi^2 R a^2 f_0 \psi_0$. Hence, $S_\theta = S_\varphi$ and, accordingly, for this solution $S = 2S_\varphi$. We shall see below that the expression $S = 2S_\varphi$ is very general). So, we obtain

$$S = -8\pi^2 R a^2 f_0 \psi_0. \qquad (4.4)$$

Express the obtained helicity (4.4) in the terms of velocity circulation $\Gamma$ on the small contour surrounding the vortex ring once, and $\Gamma_1$ on the large contour coinciding with circular directrix of the torus:

$$\Gamma = a \int_0^{2\pi} V_\theta d\theta \simeq -2\pi a^2 R \psi_0, \qquad \Gamma_1 = R \int_0^{2\pi} V_\varphi d\varphi = 2\pi f_0. \qquad (4.5)$$

For helicity in this case we get usual expression

$$S = 2\Gamma\Gamma_1. \qquad (4.6)$$

## 5. The case of non-homogeneous swirl (with maximum speed on the rotary axis).

Let us consider the case of non-homogeneous swirl. Initially, we restrict ourselves by the special case with the maximum azimuthal velocity on circular directrix of the vortex ring: $\beta = \alpha - 2 > 0$, and the swirl disappears on the surface of the ring. From the expression (3.2) for the stream function on the boundary of the torus it follows

$$\psi \simeq \frac{\alpha-1}{2}\psi_0 R^2 a^2,$$

where $a$ is the small radius of the toroidal vortex (the radius of the meridional section). Assuming that on this boundary $V_\varphi = 0$, i.e. $f\,|_{\eta=a} = 0$, we find the integration constant $f_0$ in the expression (2.11):

$$f_0 = -\sqrt{(\alpha-1)(\alpha-2)}\,\psi_0 R^2 a.$$

---

[2] $\delta(x)$ is Dirac delta function.

Inside the vortex ring $a^2 \geq z^2 + (r-R)^2$, and the azimuthal velocity is

$$V_\varphi = -\sqrt{(\alpha-1)(\alpha-2)}\,\psi_0 R^2 \frac{\sqrt{a^2-\left(z^2+(r-R)^2\right)}}{r}.$$

The azimuthal component of the vorticity is expressed in the form

$$\omega_\varphi = -\psi_0(\alpha-1)R^2 \frac{r+R}{r^2} \simeq -2\psi_0 R(\alpha-1). \tag{5.1}$$

Thus, in the considered approximation $|r-R|/R \ll 1$ for the thin vortex we have

$$V_\varphi \omega_\varphi \simeq 2\psi_0^2(\alpha-1)^{3/2}(\alpha-2)^{1/2} R^2 \sqrt{a^2-(z^2+(r-R)^2)}. \tag{5.2}$$

The integration of (5.2) by the torus volume give us the expression $S_\varphi$:

$$S_\varphi = 2\psi_0^2(\alpha-1)^{3/2}(\alpha-2)^{1/2} R^2 \cdot 2\pi R \int_0^{2\pi} d\theta \int_0^a d\eta \sqrt{a^2-\eta^2}$$

which finally has the following form

$$S_\varphi = \frac{8}{3}\pi^2 a^3 R^3 \psi_0^2 (\alpha-1)\sqrt{(\alpha-1)(\alpha-2)}. \tag{5.3}$$

To derive the expression $S_\theta$ find the corresponding velocity and vorticity components:

$$V_r \simeq -(\alpha-1)R\psi_0 \cdot z, \quad V_z \simeq (\alpha-1)R\psi_0 \cdot (r-R)$$

$$\omega_r \simeq -\psi_0 Rz \sqrt{\frac{(\alpha-1)(\alpha-2)}{a^2-[z^2+(r-R)^2]}},$$

$$\omega_z \simeq \psi_0 R(r-R)\sqrt{\frac{(\alpha-1)(\alpha-2)}{a^2-[z^2+(r-R)^2]}}.$$

Then, integrating by the volume of the ring vortex, we have

$$S_\theta = \int (V_r \omega_r + V_z \omega_z)dV = -(\alpha-1)^{3/2}(\alpha-2)^{1/2}\psi_0^2 R^2 \int \frac{z^2+(r-R)^2}{\sqrt{a^2-[z^2+(r-R)^2]}}dV.$$

Using the same substitution of variables as before is obtained

$$S_\theta = (\alpha-1)^{3/2}(\alpha-2)^{1/2} \cdot 4\pi^2 \psi_0^2 R^3 \int_0^a \frac{\eta^3}{\sqrt{a^2-\eta^2}} d\eta =$$

$$= \frac{8}{3}\pi^2 a^3 R^3 \psi_0^2 (\alpha-1)\sqrt{(\alpha-1)(\alpha-2)},$$

and, comparing this expression with (5.3) we can see that, as above, $S_\theta = S_\varphi$ and $S = 2S_\varphi$.

Express the helicity through the velocity circulation on the small ($\Gamma$) and large ($\Gamma_1$) contours (Fig. 1). For the circulation over the small contour (circle with radius $a$), using Stokes formula, we obtain:

$$\Gamma = a\int_0^{2\pi} V_\theta d\theta \simeq \pi a^2 \omega_\varphi = -2\pi a^2 R \psi_0 (\alpha-1). \tag{5.4}$$

The circulation over the large contour (circular directrix of the torus) is

$$\Gamma_1 = R\int_0^{2\pi} V_\varphi d\varphi = -2\pi R^2 a \psi_0 \sqrt{(\alpha-1)(\alpha-2)} \tag{5.5}$$

Expressing helicity through the circulations, for the case of non-homogeneous swirl with its maximum velocity on the rotary axis and zero on the boundary, the following is obtained:

$$S = \frac{4}{3}\Gamma\Gamma_1.$$

We see in the case of non-homogeneous swirl that the coefficient $k$ for the product of circulations in the expression for the helicity $S = k\Gamma\Gamma_1$ may be different from two.

## 6. Helicity of a ring vortex. General case.

Let us go back to the general representation (4.2) of helicity as the sum of two terms $S = S_\varphi + S_\theta$ – "longitudinal" $S_\varphi$ and "transversal" $S_\theta$. In the case of axial symmetry of the flow we represent the vorticity component in the form

$$\omega_\theta = \mathrm{rot}\, V_\varphi \equiv [\nabla, V_\varphi], \qquad \omega_\varphi = rot_\varphi V_\theta.$$

Consider the difference

$$S_\theta - S_\varphi = \int (V_\theta \cdot \omega_\theta - V_\varphi \cdot \omega_\varphi) dV = \int (V_\theta \cdot [\nabla, V_\varphi] - V_\varphi \cdot [\nabla, V_\theta]) dV.$$

and apply the identity $\mathrm{div}[\mathbf{a},\mathbf{b}] \equiv \nabla\cdot[\mathbf{a},\mathbf{b}] = \mathbf{b}\cdot[\nabla,\mathbf{a}] - \mathbf{a}\cdot[\nabla,\mathbf{b}]$ to the vectors. $V_\varphi$, $V_\theta$. Then we get

$$S_\theta - S_\varphi = \int \nabla \cdot \left[ \mathbf{V}_\varphi, \mathbf{V}_\theta \right] dV = \int_\Sigma \left[ \mathbf{V}_\varphi, \mathbf{V}_\theta \right] \cdot \mathbf{n} \, dS = 0,$$

if azimuthal velocity component is equal to zero on the boundary of the integration range $\Sigma$.

Thus, in the general case of axisymmetric flow in circular vortex

$$S_\theta = S_\varphi; \quad S = 2S_\varphi.. \tag{6.1}$$

At the same time the helicity expressed in terms of circulations

$$S = k\Gamma\Gamma_1 \tag{6.2}$$

for toroidal vortex is not universal: coefficient $k$ for product of circulations may be different from two. Indeed, let us consider a thin vortex ring with swirl of some general form $rV_\varphi = f(\psi) \neq const$.

In the considered solution of (3.2) $\psi = \psi(\eta)$ and $f = g(\eta)$, where $g(\eta)$ is:

$$g(\eta) \simeq \sqrt{f_0^2 - (\alpha-1)(\alpha-2)\psi_0^2 R^4 \eta^2}.$$

Then, taking into account (4.2), (5.1) and (6.1) for the helicity, the following is obtained

$$S = 2S_\varphi = -16\pi^2(\alpha-1)R\psi_0 \int_0^a \eta \cdot g(\eta) d\eta,$$

or

$$S = \Gamma\Gamma',$$

where $\Gamma$ is determined according to (5.4), and

$$\Gamma' = \frac{8\pi}{a^2} \int_0^a \eta \cdot g(\eta) d\eta.$$

If the swirl is homogeneous ($\alpha = 2$), then $g(\eta) \equiv f_0$, so $\Gamma' = 2\Gamma_1$, and, in accordance with (6.2), we get $k = 2$. For the case of non-homogeneous swirl we have $\Gamma' = k\Gamma_1$, where $k = \dfrac{4}{f_0 a^2} \int_0^a \eta \cdot g(\eta) d\eta$.

The difference from the value $k = 2$ is related to the fact that in the case of swirling vortex the linked contours belong to the same toroidal vortex and are not independent. There is still a topological integral of helicity, which can be expressed in terms of circulations of two linked contours. But here coefficient $k$ is a functional of distribution of swirl over the cross-section of the torus, and also depends on the choice of contours by which the velocity circulation is calculated.

## 7. Non-homogeneous swirl (with maximum speed on the vortex surface)

Now we will consider the case of negative $\beta = \alpha - 2 < 0$ (with $\alpha > 1$). It corresponds to the swirl, where the maximum azimuthal velocity component is achieved on the surface of ring vortex. The solution of equation (2.7) is still determined by the expression (3.2), and the value of velocity circulation $\Gamma$ on the small circuit coincides with the expression (5.4). For swirl velocity inside thin ring we obtain the expression

$$rV_\varphi = f \simeq \sqrt{f_0^2 + (\alpha-1)(2-\alpha)\psi_0^2 R^4 [z^2 + (r-R)^2]} \ .$$

Hence

$$S_\varphi = \int V_\varphi \omega_\varphi dV \simeq -8\pi^2 \psi_0 R(\alpha-1) \int_0^a \eta \sqrt{f_0^2 + (\alpha-1)(2-\alpha)\psi_0^2 R^4 \eta^2} \, d\eta \ ;$$

$$S = 2S_\varphi \simeq -\frac{16\pi^2 ((f_0^2 + (\alpha-1)(2-\alpha)\psi_0^2 R^4 a^2)^{3/2} - f_0^3)}{3\psi_0 R^3 (2-\alpha)} \ .$$

Just as in the above case of homogeneous swirl, at the torus boundary $\eta = a$ we have a vortex layer: the azimuthal velocity component at the exit from the ring is reduced to zero abruptly (by assumption, there is no swirl outside the ring). Circulation on the circular directrix $r = R$, $z = 0$ is $\Gamma_1 = 2\pi f_0$. Respectively,

$$k \equiv \frac{2S_\varphi}{\Gamma \cdot \Gamma_1} \simeq \frac{16\pi^2 \left\{[f_0^2 + (\alpha-1)(2-\alpha)\psi_0^2 R^4 a^2]^{3/2} - f_0^3\right\}}{2\pi f_0 \cdot 2\pi a^2 R \psi_0 (\alpha-1) \cdot 3\psi_0 R^3 (2-\alpha)} = \tag{7.1}$$
$$= \frac{4\left\{[f_0^2 + (\alpha-1)(2-\alpha)\psi_0^2 R^4 a^2]^{3/2} - f_0^3\right\}}{3f_0 \cdot a^2 (\alpha-1)(2-\alpha) \cdot \psi_0^2 R^4} \ .$$

Suppose now that $f_0 = f(0) = 0$. Then

$$f = \sqrt{2\psi_0(2-\alpha)} \cdot \psi R \ ,$$

or, by virtue of (3.2)

$$f \simeq -\sqrt{(2-\alpha)(\alpha-1)} \cdot \sqrt{z^2 + (r-R)^2} \psi_0 R^2 \ .$$

In this case, the swirl velocity inside the ring is directly proportional to the distance to circular directrix of the torus

$$V_\varphi = \frac{f}{r} \simeq -\sqrt{(2-\alpha)(\alpha-1)\left(z^2 + (r-R)^2\right)} \psi_0 R \ .$$

The azimuthal velocity component becomes zero on circular directrix of the ring vortex and velocity circulation on this directrix $\Gamma_1 = 0$. It is clear that in this situation $k \to \infty$ that also follows from

(7.1) when $f_0 \to 0$. Let's determine the value of helicity when $f_0 = 0$. The expression for the azimuthal component of vorticity $\omega_\varphi \simeq -2\psi_0 R(\alpha-1)$ coincides with (5.1) and, accordingly, the helicity is equal to

$$S_\varphi = \int V_\varphi \omega_\varphi dV \simeq 2\int \psi_0^2 R^2 (\alpha-1)^{3/2} \sqrt{(2-\alpha)(\alpha-1)[z^2+(r-R)^2]} dV.$$

$$S = 2S_\varphi = \frac{16}{3}\pi^2(\alpha-1)\sqrt{(\alpha-1)(2-\alpha)} \cdot a^3 R^3 \psi_0^2$$

## 8. Discussion

Below we will explain the above results on the example of non-homogeneous swirl with maximum velocity on the torus circular axis (see section 4).

The value of helicity integral for the swirling vortex ring with specified circulation velocity on the small $(\Gamma)$ and large $(\Gamma_1)$ contours depends on distribution of the swirl along small radius of the torus. At the same time, swirl distribution is unambiguously associated with distribution of the vorticity meridional component.

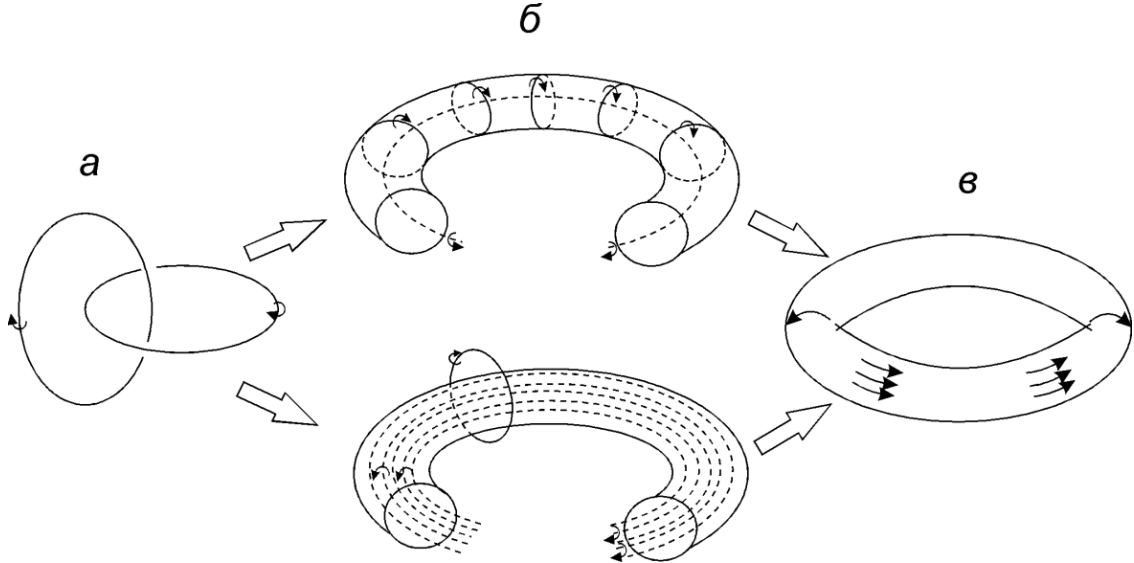

**Fig. 2. a) Two linked vortex threads.** Small circular arrows show the direction of circular motion near each of the threads. *(In the remaining pictures the meaning of circular arrows is the same.)* **b) The upper picture.** Circles in the meridional plane represent the family of vortex threads (vortex shroud) obtained as a result of 'smearing' of vertical vortex thread over the torus surface (the meaning of circular arrows - see above). **The lower picture.** Circles in horizontal planes represent the family of vortex threads obtained as a result of 'smearing' of horizontal vortex thread over the torus volume. **c) Vortex ring with homogeneous swirl.** Here the circular arrows show the direction of movement in the meridional and horizontal sections (in the latter case, the arrows have the same length, since the swirl is homogeneous).

If to replace the continuous vorticity distribution with discrete one, separating the azimuthal component from the poloidal (meridian) component without changing the helicity value, we will get two families of linked vortex threads.

Integral of the helicity for the system of two linked vortex threads is determined by the product of circulations along a contour enclosing these threads: $S = 2\Gamma\Gamma_1$ (Fig. 2a). As to continuous distribution, let us 'smear' one of the vortex threads over the surface of circular torus, preserving the meridionality of vorticity $\omega_\varphi = 0$, $\omega_\theta \neq 0$ (Fig. 2b, upper picture), and the other thread – over the inside of the same torus (2b, lower picture), so that the vorticity was oriented along the azimuth $\omega_\varphi \neq 0$, $\omega_\theta = 0$ (preserving the value of circulations $\Gamma, \Gamma_1$). At that the value of helicity integral will not change.

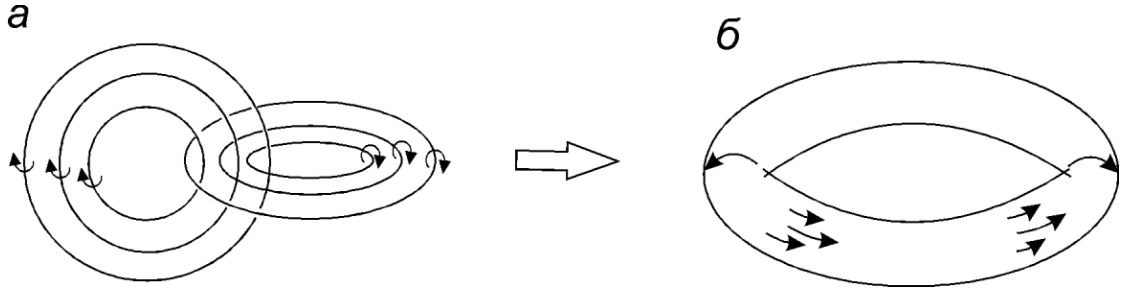

**Fig. 3.** a) **Two families of linked vortex threads** with incomplete links of vertical and horizontal threads. b) **The ring vortex with inhomogeneous swirl.** The circular arrows show the direction of movement in meridional and horizontal sections (in the latter case, the arrows are of different lengths, since the swirl is non-homogeneous).

If to 'smear' the thread with meridional vorticity over the torus volume (not the surface), the result will be different: not all pairs of partial threads, belonging to different families, will be linked. This is the reason why the coefficient k in (6.2) may be less than 2.

## 9. Conclusion

For a vortex with a swirl (orbital motion) the helicity is nonzero, but relation to the product of the linked contours circulations differs from the known formula $S = 2\Gamma\Gamma_1 \cdot I$, where factor $I$ is the Gaussian integral of links with integer values [8, 12, 13]. In our case, for thin vortex rings with circular cross-section the coefficient $k$ in (6.2) may vary in the limits $4/3 \leq k < \infty$, if the values $\Gamma$ and $\Gamma_1$ of velocity circulation are determined on the above-mentioned circuits (small and large generatrices of the torus). Graphic explanation of this difference is presented in Figures 2 and 3. The case when $k = 2$ corresponds to the homogeneous swirl, and the cases when $4/3 \leq k < 2$ and

$2 < k < \infty$ correspond to the non-homogeneous swirl, with maximum and minimum of azimuthal velocity on circular directrix of the torus. The minimum possible value $k = 4/3$ corresponds to the considered parabolic case of distribution of azimuthal velocity (2.8) in the absence of swirl on the vortex boundary. [3]

# References

**1. P.G. Saffman.** *Vortex Dynamics* (Cambridge University Press, Cambridge, 1995; Nauchnyi Mir, Moscow, 2000).

**2. G.V. Levina and M.T. Montgomery.** On the first investigation of the spiral structure of tropical cyclogenesis. *Dokl. Earth Sci.* **434** *(Part 1), 1285 (2010).*

**3. V. Bjerknes.** Solar hydrodynamics. *Astrophys.J.* **6**, *p. 93-121, (1926)*

**4. E.Yu. Bannikova, V.M. Kontorovich and S.A. Poslavsky.** Influence of Orbital Motion on the Collapse of Ring Vortices in an Accretion Flow. *J. Exp. Theor. Phys.* **119** *(3), 584–589 (2014).*

**5. H.K. Moffatt & A. Tsinober.** Helicity in laminar and turbulent flow. *Annu. Rev. Fluid Mech.. **24**: 281-312 (1992)*

**6. Takashi Naitoh, Nobuyuki Okura, Toshiyuki Gotoh, and Yusuke Kato**. On the evolution of vortex rings with swirl. *Physics of fluids **26**, 067101 (2014)*

**7**. **F.V. Dolzhanski**. *Lectures on Geophysical Hydrodynamics*. M.:ICM RAN (2006)

**8. A.V. Tur and V.V. Yanovskii.** *Hydrodynamic Vortex Structures* (State Scientific Institution "Institute for Monocrystals" of the National Academy of Sciences of Ukraine, Kharkiv, 2012) [in Russian].

**9. Yu.P. Ladikov-Roev and O.K. Cheremnykh.** *Mathematical Models of Continuous Media* (Naukova Dumka, Kiev, 2010) [in Russian]

**10. L.D. Landau and E.M. Lifshitz.** *Course of Theoretical Physics: Volume 8. Electrodynamics of Continuous Media* (Nauka, Moscow, 1982; Butterworth–Heinemann, Oxford, 1984), p. 164.

**11. V.I. Petviashvili and O.A. Pokhotelov.** *Solitary Waves in Plasmas and in the Atmosphere* (Energoatomizdat, Moscow, 1989; Gordon and Breach, London, 1982).

**12. B.A. Dubrovin, S.P. Novikov and A.T. Fomenko**. *Modern Geometry. Methods and Applications.*— Springer-Verlag, GTM 93, Part 1, 1984; GTM 104, Part 2, 1985. Part 3, 1990, GTM 124.

**13**. **V.I. Arnol'd, B.A. Khesin.** Topological methods in hydrodynamics. *Annual Review of Fluid Mechanics. Vol. 24 (A92-45082 19-34). Palo Alto, CA, Annual Reviews, Inc., 1992, p. 145-166.*

**14. E.Yu. Bannikova, V.M. Kontorovich and S.A. Poslavskii.** Helicity of the toroidal vortex with swirl. Problems of Atomic Science and Technology, №4 (98), p. 144-147 (2015).

---

[3] The materials of article were reported on the XIII International conference "Plasma electronics and new acceleration methods" (Kharkov, Ukraine, 2015) and partly published in [14].